\documentclass[twocolumn, tighten]{aastex62}

\usepackage[utf8]{inputenc}
\usepackage{tgtermes}
\shortauthors{\footnotesize{Hu et al.}}
\received{\footnotesize{2018 March 31}}
\revised{\footnotesize{2018 July 19}}
\accepted{\footnotesize{2018 July 22}}

\usepackage{color}
\usepackage{amsmath}
\usepackage{natbib}
\hypersetup{breaklinks=true, colorlinks=true, urlcolor={blue}, citecolor={blue}, linkcolor={blue}}

\usepackage{etoolbox}

\makeatletter
\patchcmd{\NAT@citex}
  {\@citea\NAT@hyper@{%
     \NAT@nmfmt{\NAT@nm}%
     \hyper@natlinkbreak{\NAT@aysep\NAT@spacechar}{\@citeb\@extra@b@citeb}%
     \NAT@date}}
  {\@citea\NAT@nmfmt{\NAT@nm}%
   \NAT@aysep\NAT@spacechar\NAT@hyper@{\NAT@date}}{}{}

\patchcmd{\NAT@citex}
  {\@citea\NAT@hyper@{%
     \NAT@nmfmt{\NAT@nm}%
     \hyper@natlinkbreak{\NAT@spacechar\NAT@@open\if*#1*\else#1\NAT@spacechar\fi}%
       {\@citeb\@extra@b@citeb}%
     \NAT@date}}
  {\@citea\NAT@nmfmt{\NAT@nm}%
   \NAT@spacechar\NAT@@open\if*#1*\else#1\NAT@spacechar\fi\NAT@hyper@{\NAT@date}}
  {}{}%

\makeatother

\newcommand{\chandra}{\emph{Chandra}}

\newcommand{\rosat}{\emph{ROSAT}}

\newcommand{\xmm}{\emph{XMM-Newton}}
\newcommand{\swift}{\emph{Swift}}

\newcommand{\nustar}{\emph{NuSTAR}}

\newcommand{\ms}{$M_{\odot}$}

\newcommand{\lumcgs}{erg~s$^{-1}$}

\begin{document}
\title{NGC 7793 P9: An Ultraluminous X-Ray Source Evolved from a Canonical Black Hole X-Ray Binary}
\author{Chin-Ping Hu}
\affiliation{\emph{Department of Physics, The University of Hong Kong, Pokfulam Road, Hong Kong;} \href{mailto:cphu@hku.hk}{\emph{cphu@hku.hk}}}
\author{Albert K. H. Kong}
\affiliation{\emph{Institute of Astronomy, National Tsing Hua University, Hsinchu 30013, Taiwan}}
\author{C.-Y. Ng}
\affiliation{\emph{Department of Physics, The University of Hong Kong, Pokfulam Road, Hong Kong;} \href{mailto:cphu@hku.hk}{\emph{cphu@hku.hk}}}
\author{K. L. Li}
\affiliation{\emph{Department of Physics and Astronomy, Michigan State University, East Lansing, MI 48824, USA}}



\begin{abstract}
Transient ultraluminous X-ray sources (ULXs) provide an important link bridging transient low-mass X-ray binaries and ULXs. Here we report the first discovery of both a canonical sub-Eddington outburst and an ultraluminous super-Eddington outburst from an unusual transient ULX, NGC~7793~P9 with a variability factor higher than $10^3$. Its X-ray spectrum switches between the typical high/soft state and the steep power-law state during the canonical outburst. The inner radius of the accretion disk and the disk temperature--luminosity correlation suggest that P9 harbors a stellar-mass black hole (BH). At the beginning of the ultraluminous outburst, we observe a cool outer disk with a hard Comptonized spectrum, implying a transition to the ULX regime. When the luminosity increases to $L\gtrsim3\times10^{39}$\,\lumcgs, P9 shows a significantly curved spectrum that can be described by either a slim disk or a strongly curved Comptonized corona. The phenomenological model suggests that the hot disk observed near the peak of the ultraluminous outburst is coincidentally consistent with the extension of the thermal track. Utilizing more physical Comptonized disk models, we suggest that the corona cools down and the apparent disk-like spectrum is dominated by soft Comptonization. The significant variability above 1\,keV supports this two-component scenario. The spectral evolution can also be interpreted with the supercritical accretion model. All these indicate that a canonical black hole X-ray binary can show properties of a ULX.
\end{abstract}

\keywords{accretion, accretion disks -- black hole physics -- X-rays: binaries -- X-rays: individual (NGC~7793~P9)}

\section{Introduction}\label{introduction}
Accreting black holes (BHs) are bright X-ray sources powered by the gravitational potential energy of accreting materials. The luminosities of Galactic BHs are strongly limited to the Eddington luminosity ($L_{\rm{Edd}}$), at which the radiation pressure balances the inward gravitational force of the accreting materials. Thanks to the revolution in astronomical instrumentation over the past several decades, the discovery of ultraluminous X-ray sources (ULXs) challenged the Eddington limit \citep[see reviews by][]{FrankKR2002, FengS2011, KaaretFR2017}. ULXs are extragalactic off-nuclear X-ray point sources with luminosities exceeding the Eddington limit of a stellar-mass BH. Most ULXs are believed to be powered by stellar-mass BHs with mild beaming and/or moderate super-Eddington accretion rates. They show ultraluminous states distinct from the canonical spectral states of BH X-ray binaries \citep[BHXBs, ][]{FengS2011, GladstoneRD2009, KaaretFR2017}. Unlike the spectral behavior of a BHXB in the steep power-law (SPL) state that shows a hot disk and an optically thin corona, a genuine ULX usually shows a much cooler disk with a larger inner radius plus an optically thick corona and has a luminosity of $L\gtrsim3\times10^{39}$\,\lumcgs. ULX states can be further divided into the hard ultraluminous (HUL), soft ultraluminous (SUL), and supersoft ultraluminous spectral states \citep{SuttonRM2013, KaaretFR2017}. In addition, those low-luminosity ULXs with $L\lesssim3\times10^{39}$\,\lumcgs usually show a curved spectrum that can be described with a broadened disk (BD) model, implying that the disk component is no longer a geometrically thin accretion disk but the luminosity remains sub-Eddington \citep{SuttonRM2013}. 

The best studied ULXs are persistent sources with low variability factors \citep{FengK2006,KaaretF2009}. A stable high mass accretion rate is necessary to interpret the long lifetime of the ultraluminous state. As long as the companion star of a ULX is a massive star with $M\gtrsim10M_\odot$, super-Eddington mass transfer occurs when the companion evolves to the Hertzsprung gap without a common envelope \citep{KingDW2001, RappaportPP2005, PavlovskiiIB2017}. Most of the currently known ULXs, including the ultraluminous pulsar P13 in NGC 7793, have early-type companions with high optical luminosities \citep{BachettiHW2014, MotchPS2014, LauKB2016}. It has also been proposed that some ULXs may exhibit a low-mass companion and be powered by a long-lasting super-Eddington outburst \citep{King2002}. In contrast, BHXBs with low-mass companions are usually transient X-ray sources. Therefore, transient ULXs is key to testing this scenario and showing that a low-mass X-ray binary (LMXB) can evolve to a ULX with typical UL spectral behaviors. The Galactic LMXBs GRS 1915+105 and XTE J1550$-$564 showed optically thick Comptonized spectra during the peak of the outburst with a luminosity of $\sim10^{39}$\,\lumcgs. They could fill in the link bridging BHXBs and ULXs \citep{KubotaD2004, Soria2007, Vierdayanti2010}.  Several transient ULXs were reported as evolved from canonical BHXBs although they showed either canonical BHXB spectral states or ultraluminous outburst only \citep[see, e.g., ][]{MiddletonSR2012, KaaretF2013}.  Therefore, it remains unclear whether a canonical BHXB showing sub-Eddington outbursts can evolve into a ULX with a much higher luminosity and typical UL spectral behaviors. In this paper, we report the discovery of a transient ULX, NGC~7793~P9 (hereafter P9), that showed both a canonical BHXB outburst and an ultraluminous outburst that fill in the gap. 

NGC~7793 is an SA(s)d type galaxy in the Sculptor Group with a distance of 3.6--3.9\,Mpc \citep{KarachentsevGS2003, Crowther2013, Radburn-SmithJS2011, TullyCS2016}. It has a bulge with filamentary spiral structures and several H{\sc ii} regions \citep{ReadP1999}. P9 is located at the northeast edge of one filament extending from the host galaxy and has no obvious association with any H{\sc ii} regions or supernova remnants \citep{ReadP1999}. P9 was first characterized as a highly variable point source in a \rosat\ survey \citep{ReadP1999}. The X-ray spectrum of P9 was well fit with either a PL or a thermal bremsstrahlung model. The absorption column density ($N_{\textrm{H}}$) was rather large. A deep \chandra\ observation (ObsID \dataset[3954]{http://cda.harvard.edu/chaser/viewerContents.do?obsid=3954}) revealed a faint source, CXOU J235808.7$-$323403, at the position of P9, but its spectrum was not well characterized \citep{PannutiSF2011}. Therefore, the luminosity was not well constrained. By simply assuming a PL with a photon index $\Gamma=1.5$ and $N_{\textrm{H}}=1.15\times10^{20}$\,cm$^{-2}$, the luminosity of P9 was estimated to be $4.5 \times10^{37}$\,\lumcgs. No optical or radio counterparts have been found \citep{ReadP1999, PannutiSF2011}.

We report a detailed analysis of the transient ULX P9 based on the \emph{Neil Gehrels Swift} (hereafter \swift) monitoring data and several dedicated observations made with \chandra, \xmm, and \nustar. We introduce the observation and data analysis in Section \ref{observation}. The detailed spectral analysis, including the variability of $\Gamma$, the stacked \swift\ spectroscopy, and tests of more physical models during the ultraluminous outburst are presented in Section \ref{result}. We further discuss the implications of the spectral variability, especially the evolution of the disk component and the accretion models, in the super-Eddington accretion regime in Section \ref{discussion}. Finally, we summarize this work in Section \ref{summary}.

\section{Observations and Data Reduction}\label{observation}

\begin{figure}
\includegraphics[width=0.47\textwidth]{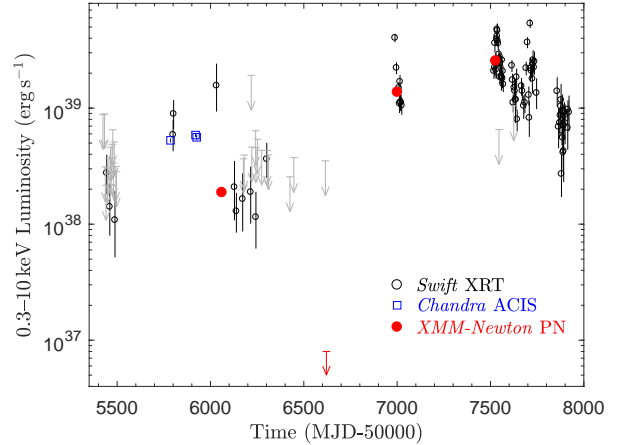} 
\caption{The \swift\ XRT light curve of NGC~7793~P9. The black diamonds are the data points determined with \swift\ XRT, where the gray crosses denote the upper limits. The \chandra\ (blue squares) and \xmm\ (red filled circles) data are also plotted on the same figure.}
\label{p9_lc}
\end{figure}

\begin{deluxetable*}{cccccc}
\tabletypesize{\footnotesize}
\tablecaption{Observation Log of NGC 7793 P9.} 
\label{observation_log}
\tablehead{\colhead{Epoch} & \colhead{Observatory} & \colhead{ObsID} & \colhead{Instrument} & \colhead{MJD} & \colhead{Exposure$^a$ (ks)}}
\startdata
C1 & \chandra\ & \dataset[3954]{http://cda.harvard.edu/chaser/viewerContents.do?obsid=3954} & ACIS-S & 52889 & 49 \\
C2 & \chandra\ & \dataset[14231]{http://cda.harvard.edu/chaser/viewerContents.do?obsid=14231} & ACIS-S & 55787 & 59\\
C3 & \chandra\ & \dataset[13439]{http://cda.harvard.edu/chaser/viewerContents.do?obsid=13439} & ACIS-S & 55920 & 58\\
C4 & \chandra\ & \dataset[14378]{http://cda.harvard.edu/chaser/viewerContents.do?obsid=14378} & ACIS-S & 55926 & 25\\
X1 & \xmm\ & \dataset[0693760101]{http://nxsa.esac.esa.int/nxsa-web/\#obsid=0693760101} & PN & 56061 & 23 \\
 & & & MOS & & 36 \\
X2 & \xmm\ & \dataset[0693760401]{http://nxsa.esac.esa.int/nxsa-web/\#obsid=0693760401} & PN & 56621 & 46 \\
 & & & MOS & & 48 \\
X3 & \xmm\ & \dataset[0748390901]{http://nxsa.esac.esa.int/nxsa-web/\#obsid=0748390901} & PN & 57002 & 47\\
 & & & MOS & & 49 \\
X4$^b$ & \xmm\ & \dataset[0781800101]{http://nxsa.esac.esa.int/nxsa-web/\#obsid=0781800101} & PN & 57528 & 24\\
 & & & MOS & & 48 \\
 & \nustar\ & 80201010002 & FPMA+FPMB & 57528 & 114
\enddata
\tablenotetext{a}{Effective exposure time after filtering out the time intervals with flaring background.}
\tablenotetext{b}{A joint observation with \nustar.}
\vspace{-0.9cm}
\end{deluxetable*}

\subsection{\emph{Swift} Observations}\label{swift_obs}
The host galaxy NGC 7793 has been monitored with \swift\ since 2010. Four series of observations have been carried out in late 2010, from mid 2011 to late 2013, in late 2014, and from 2016 April to 2017 June.  We extracted the XRT light curve with the online XRT pipeline\footnote{\url{http://www.swift.ac.uk/user_objects/}} \citep{EvansBP2007, EvansBP2009}. P9 is clearly detected in 88 of the total 122 observations. We converted the 0.3--10\,keV count rates to the luminosities with WebPIMMS\footnote{\url{https://heasarc.gsfc.nasa.gov/cgi-bin/Tools/w3pimms/w3pimms.pl}} by assuming $N_{\rm{H}}=10^{21}$\,cm$^{-2}$, a PL spectrum with $\Gamma=2$, and a distance of 3.9\,Mpc. We will show that $\Gamma$ varied between $\sim$1.5 and $\sim$2.5 in a later section. The value $\Gamma=2$ is a conventional average value to estimate the luminosity variability in Figure \ref{p9_lc}.  We searched for periodic signals between $P=10$--$1000$\,days to check if P9 contained orbital or superorbital modulations like other ultraluminous pulsars but we did not find significant periodicity \citep{KongHL2016, WaltonFB2016, HuLK2017}. The luminosity of P9 varied between $\sim10^{38}$ and $\sim10^{39}$\,\lumcgs, well below the ULX regime before 2013 (see Figure \ref{p9_lc}). Interestingly, P9 showed an ultraluminous outburst with a peak luminosity of $\sim5\times10^{39}$\lumcgs\ after 2014, and the luminosity returned to $\lesssim 10^{39}$\,\lumcgs\ in 2017.  We further downloaded all of the \swift\ UVOT images taken between 2010 and 2017, and stacked them to obtain deep UV images in four \swift\ bands (\emph{u}, \emph{w1}, \emph{m2}, and \emph{w2}). No UV counterpart was seen within the error circle of the \chandra\ position.

\subsection{\chandra\ Data}
NGC 7793 was observed with \chandra\ in 2003 (C1: ObsID \dataset[3954]{http://cda.harvard.edu/chaser/viewerContents.do?obsid=3954}) and 2011 (C2: ObsID \dataset[14231]{http://cda.harvard.edu/chaser/viewerContents.do?obsid=14231}, C3: ObsID \dataset[13439]{http://cda.harvard.edu/chaser/viewerContents.do?obsid=13439}, C4: ObsID \dataset[14378]{http://cda.harvard.edu/chaser/viewerContents.do?obsid=14378}). The observation log of P9 is listed in Table \ref{observation_log}. These four observations were made with the Advanced CCD Imaging Spectrometer array (ACIS-S) in the timed exposure mode. We reprocessed the data using the pipeline \texttt{chandra\_repro} in the \chandra\ Interactive Analysis of Observations (CIAO) version 4.9 with the most recent calibration database (CALDB) version 4.7.3 \citep{FruscioneMA2006}. We extracted the source events from a circular aperture of 2\arcsec\ radius. The background events were extracted from a nearby source-free region. We estimated the source luminosity with the \texttt{srcflux} command and plotted the results in Figure \ref{p9_lc}. The luminosities of the 2011 observations are $\sim5\times10^{38}$\,\lumcgs, consistent with the \swift\ light curve. 

We extracted the light curves of P9 to check for any variability in each observation. We used the task \texttt{dmextract} to create the background-subtracted light curves, and the results are plotted in Figure \ref{chandra_xmm_lightcurve}. The light curves show limited fluctuations, and no significant trends were observed.

\subsection{\xmm\ Data}
\xmm\ has observed NGC 7793 nine times with the EPIC CCD in the full-frame mode, and four of them are publicly available. We reduced the data using the \xmm\ Science Analysis Software (SAS version 16.0.0) and used \texttt{epproc} and \texttt{emproc} to reprocess the PN and MOS data with the latest calibration files, respectively. We filtered out the time intervals with a flaring particle background. The effective exposure times are listed in Table \ref{observation_log}. The 2012 observation (X1: ObsID \dataset[0693760101]{http://nxsa.esac.esa.int/nxsa-web/\#obsid=0693760101}) shows that the luminosity of P9 was $\sim2\times10^{38}$\,\lumcgs. In 2013, P9 was not detected with \xmm\ (X2: ObsID \dataset[0693760401]{http://nxsa.esac.esa.int/nxsa-web/\#obsid=0693760401}). We estimated a 3$\sigma$ upper limit as $\sim8\times10^{36}$\,\lumcgs. On the other hand, P9 was very bright with luminosities exceeding $10^{39}$\,\lumcgs\ in 2014 (X3: ObsID \dataset[0748390901]{http://nxsa.esac.esa.int/nxsa-web/\#obsid=0748390901}) and 2016 (X4: ObsID \dataset[0781800101]{http://nxsa.esac.esa.int/nxsa-web/\#obsid=0781800101}). The nondetection of P9 in 2013 suggests that P9 was in quiescence, and the entire time span can be divided into two epochs: a less luminous outburst from 2011 to 2012 and an ultraluminous outburst from 2014. 

We then extracted the background-subtracted light curves of P9 from individual observations. We extracted the source events from a circular aperture of 15\arcsec\ radius to minimize the possible contamination from CXOU J235810.4$-$323357 \citep{PannutiSF2011}. We used \texttt{epiclccorr} to subtract the background light curve and apply the relative and absolute corrections.  Figure \ref{chandra_xmm_lightcurve} shows the light curves of the \xmm\ observations. No significant variability was seen in the X1 and X3 data sets. However, the light curve of X4 shows a decreasing trend. We calculated the fractional variability ($F_{\rm{var}}$) for X3 and X4 in which P9 was bright enough. $F_{\rm{var}}$ can be defined as
\begin{equation}
F_{\rm{var}}=\frac{1}{\left< X \right>}\sqrt{S^2-\left< \sigma_{err}^2 \right>}\rm{,}
\end{equation}
where $S^2$ is the total variance of the light curve, $\left< \sigma_{err}^2 \right>$ is the mean square of the uncertainty of each data point, and $\left< X \right>$ is the mean count rate \citep{EdelsonTP2002}. For individual data sets, we calculated $F_{\rm{var}}$ for the full energy range (0.3--10\,keV), the soft X-ray band (0.3--1\,keV), and the hard X-ray band (1--10\,keV). We did not observe significant variability in all the three bands of X3, and the 2$\sigma$ upper limit is estimated as $F_{\rm{var}}<0.06$. On the other hand, X4 shows a clear variability with $F_{\rm{var}}=0.12\pm0.01$ in 0.3--10\,keV, and the variability is dominated by the X-ray emission in 1--10\,keV with $F_{\rm{var}}=0.15\pm0.02$. $F_{\rm{var}}$ in the 0.3--1\,keV band was less obvious with a value of $0.05\pm0.02$. We corrected the photon arrival time to the solar barycenter and searched for periodicity in the range of 0.1--1000\,s with both the $H$-test \citep{deJagerRS1989} and the Fourier analysis, but no periodicity was found.

\subsection{\nustar\ Data}
\begin{figure*}
\includegraphics[width=0.95\textwidth]{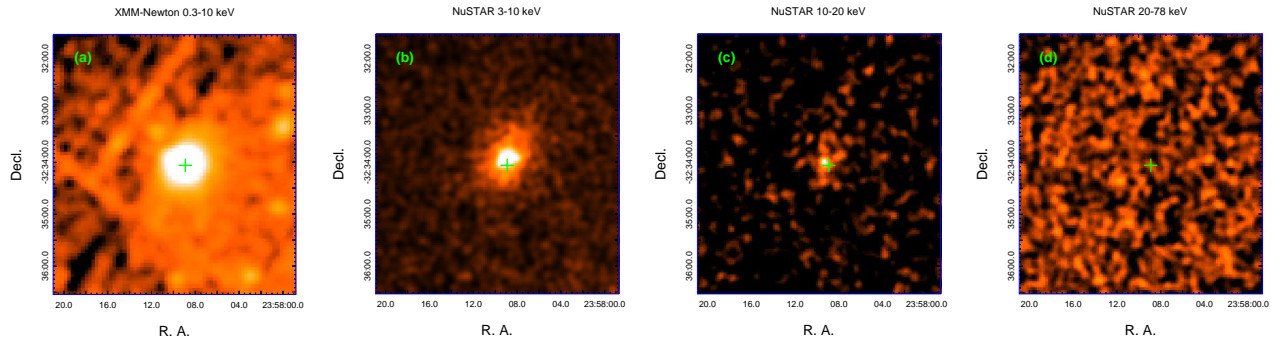} 
\caption{X-ray images of NGC~7793~P9 obtained with (a) \xmm\ in 0.3--10\,keV, and \nustar\ in (b) 3--10\,keV,(c) 10--20\,keV, and (d) 10--79\,keV. The \xmm\ image is stacked from MOS1, MOS2, and PN images, and the \nustar\ images are stacked from the FPMA and FPMB images. The green cross denotes the \chandra\ position of P9.}
\label{xmm_nustar_image}
\end{figure*}

NGC 7793 was observed with both \nustar\ and \xmm\ on 2016 May 20 (ObsID 80201010002). We reduced the data using the \nustar\ analysis software NUSTARDAS (version 1.9.1) with the calibration database of version 20170614. We used the script \texttt{nupipeline} to apply data calibration and screening, and then extracted the images, light curves, and spectra using \texttt{nuproducts}. The X-ray emission of P9 was detected with both the FPMA and FPMB below $\sim$20\,keV.  Figure \ref{xmm_nustar_image} shows the stacked images obtained by all of the \xmm\ EPIC detectors in the 0.3--10\,keV band, and \nustar\ FPMA+FPMB in the 3--10\,keV, 10--20\,keV, and 20--79\,keV ranges. The emission of P9 is dominated by soft X-rays below 10\,keV. For the 10--20\,keV band, P9 was marginally detected with a detection significance of 8.8$\sigma$ using the wavelet source detection algorithm \texttt{wavdetect}.

\begin{figure*}
\begin{minipage}{0.49\linewidth}
\includegraphics[width=0.95\textwidth]{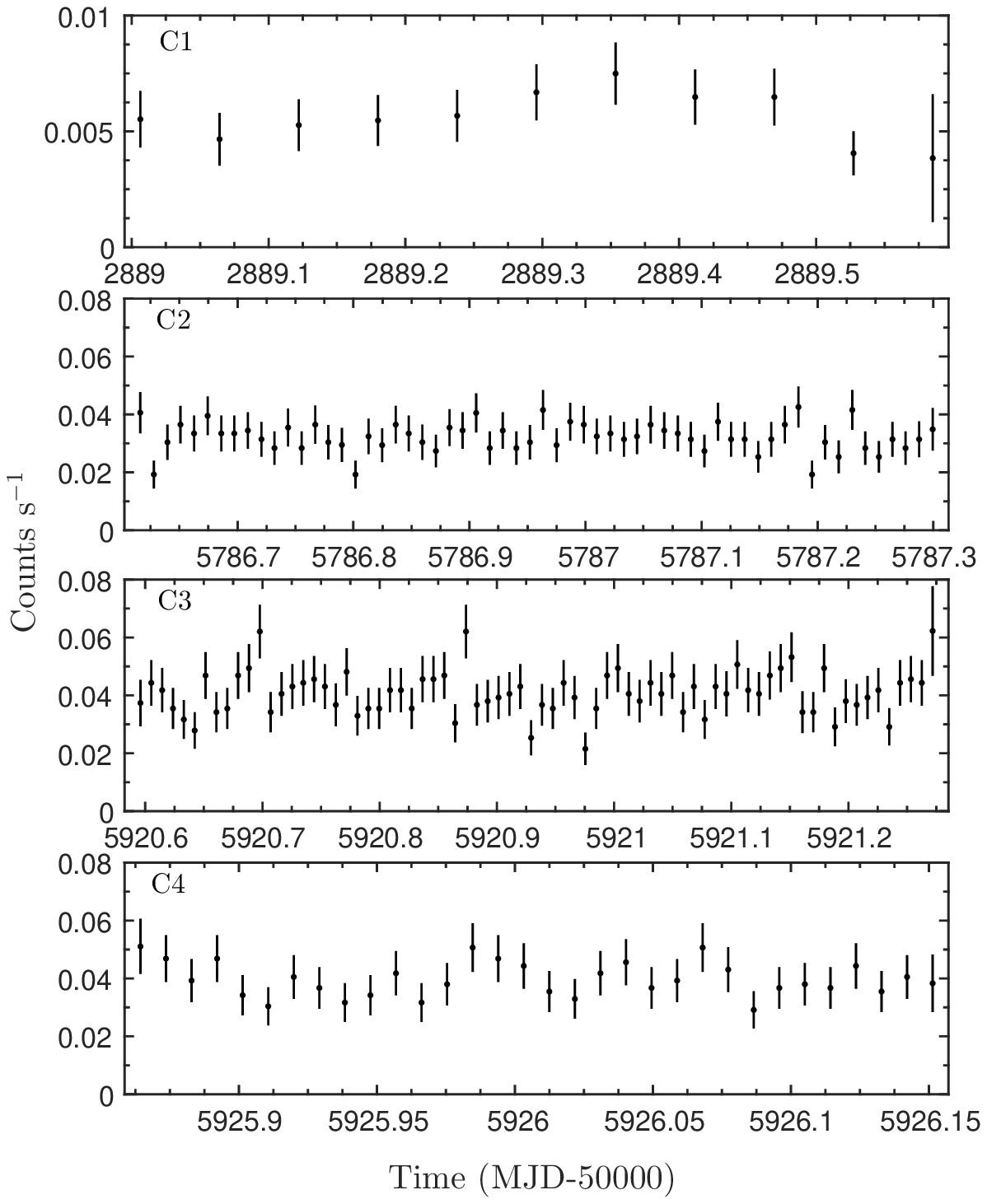}
\end{minipage}
\begin{minipage}{0.49\linewidth}
\includegraphics[width=0.95\textwidth]{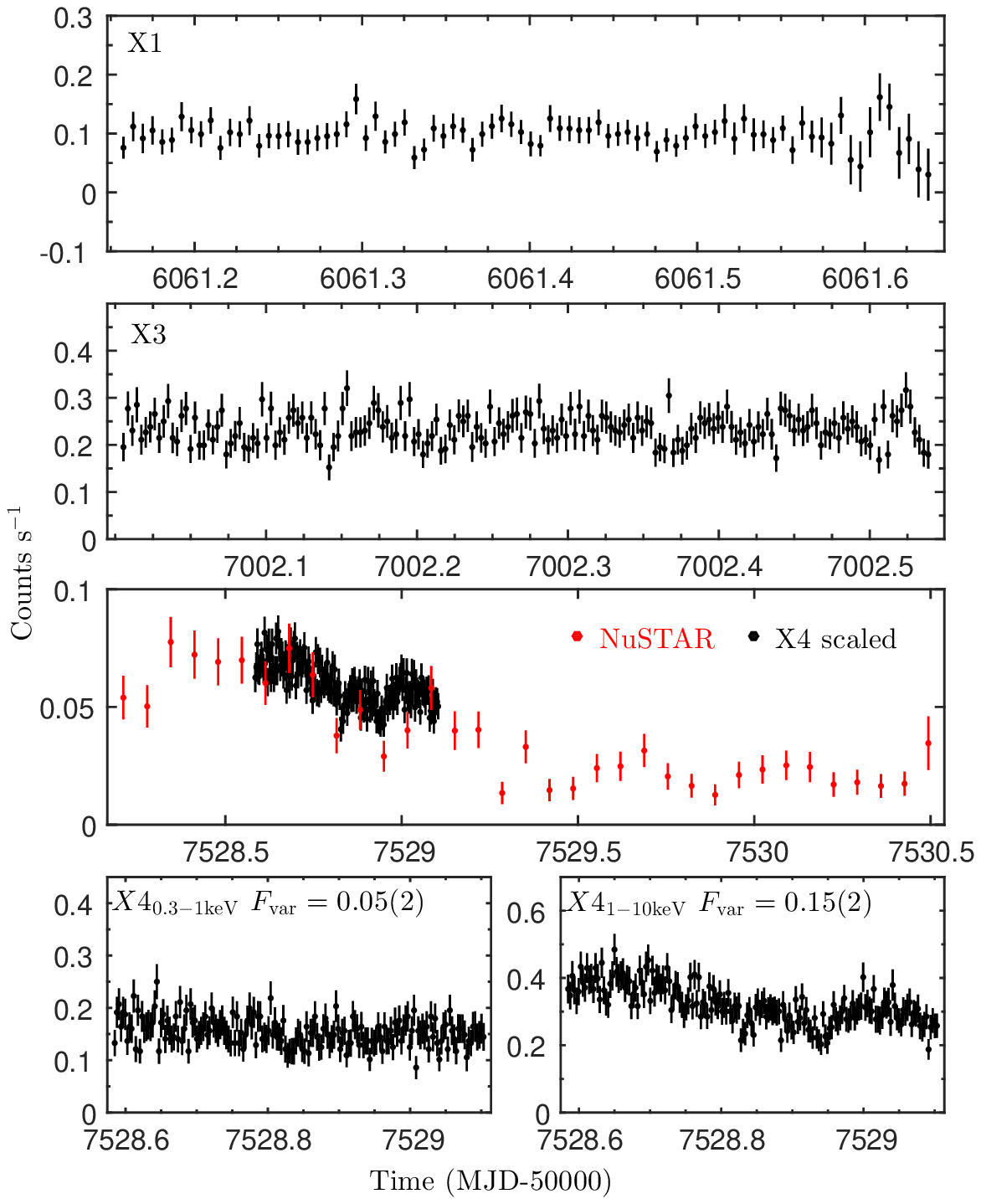}
\end{minipage}
\caption{ (Left) \chandra\ 0.5--7\,keV light curves of NGC~7793~P9. (Right) \xmm\ (0.3--10\,keV) and \nustar\ (3--79\,keV) light curves of NGC~7793~P9. The X4 light curve on the third panel was scaled to fit the \nustar\ count rate. The bottom panel showed the soft (0.1--1\,keV) and hard (1--10\,keV) X-ray light curves of X4 observation.}
\label{chandra_xmm_lightcurve}
\end{figure*}

We extracted the light curve of each \nustar\ orbit to check if P9 showed long-term variability.  We extracted the source events from a circular aperture of 60\arcsec\ radius and background events from an annulus with an inner radius of 80\arcsec\ and an outer radius of 120\arcsec. The result is plotted in Figure \ref{chandra_xmm_lightcurve} together with the scaled X4 light curve obtained with \xmm. The \nustar\ flux dropped significantly after $\sim$MJD 57529.5. 

\section{X-Ray Spectroscopy}\label{result}
We started the spectral analysis with the \chandra, \xmm, and \nustar\ data sets, which have much better signal-to-noise ratios than the \swift\ data sets. We used the absorption model {\tt tbnew}\footnote{\url{http://pulsar.sternwarte.uni-erlangen.de/wilms/research/tbabs/}} \citep{WilmsAM2000}, fixed the Galactic absorption column density to $N_{\rm{H}}(\rm{gal})=1.23\times10^{20}$\,cm$^{-2}$, and set an additional absorption for the external galaxy \citep{Kalberla2005}. We followed the empirical classification strategy of ULX spectral states proposed by \citet{SuttonRM2013}. We first employed simple phenomenological models, the multicolor disk blackbody \citep[MCD; ][]{MitsudaIK1984} and the power-law (PL), to fit the spectra. Then, we tried a combination of them (MCD+PL) to check if adding an additional component improves the fit.  For this case, we further check the empirical classification of the UL spectra by computing the photon index ($\Gamma$) and the fraction of the PL flux in 0.3--1\,keV. After that, we fit the spectrum with more physical models like the slim disk and the Comptonized models to further characterize the spectral behaviors of those data sets with the UL spectral features. The best-fit parameters for each \chandra\ and \xmm\ observations with a simple MCD+PL model were summarized in Table~\ref{xmm_chandra_parameter}.

\subsection{Phenomenological Models}
\subsubsection{\chandra\ Spectroscopy}\label{chandra_spec}
The \chandra\ 0.5--7\,keV spectra were generated using the \texttt{specextract} command, which creates source and background spectra, corresponding response files (rmf), and ancillary files (arf). The spectra were then grouped to a minimum of 15 photons per bin for C1 and 25 photons per bin for C2--C4 for spectral fitting. The C1 observation made in 2003 can be well described with a simple PL but $N_{\rm{H}}=1.3^{+1.7}_{-1.3}\times10^{21}$\,cm$^{-2}$ is not well constrained. The uncertainties of the spectral parameters are obtained within a 90\% confidence interval. The photon index is $\Gamma=2.4^{+0.5}_{-0.4}$, which is quite soft but with a large uncertainty. To doubly confirm the result obtained with the $\chi^2$ statistic, we tried the Cash statistic without grouping the photons and yielded a consistent result \citep{Cash1979}. The C2 spectrum is well fit by the MCD with an inner-disk temperature $kT_{\rm{in}}=0.96_{-0.06}^{+0.07}$\,keV. The normalization constant indicates an apparent inner radius of $R_{in}\sqrt{\cos{\theta}}\approx50$\,km where $\theta$ is the inclination angle. We tried the MCD+PL model but obtained no significant improvement. The C3 and C4 observations were taken four months after the C2 observation and had a time separation of five days. An absorbed PL model with $\Gamma\approx2.3$ and $N_{\rm{H}}\sim10^{21}$\,cm$^{-2}$ can well describe both spectra, and adding an MCD does not improve the fit statistic significantly. All of the \chandra\ spectra and the corresponding best-fit models are shown in Figure \ref{chandra_xmm_spec}. It is clear that the C2 spectrum, which was best described with an MCD model, has a stronger curvature than other three data sets that can be described by a PL model. We checked the spectral fits with Cash statistics and the results are fully consistent with those obtained with $\chi^2$ statistics. 

\subsubsection{\xmm\ Spectroscopy}\label{xmm_spec}
To extract high-quality \xmm\ spectra, we filtered out the background flaring epochs and chose the events with good grades by restricting \texttt{FLAG$=$0}. We set \texttt{PATTERN$\leq$4} for the PN data and \texttt{PATTERN$\leq$12} for the MOS data. The source and background regions were normalized using the \texttt{backscale} command, and the corresponding response files and ancillary files were created using \texttt{rmfgen} and \texttt{arfgen}. The spectra were then grouped to a minimum of 15 photons per bin for X1 and 25 photons per bin for the X3 and X4 observations using \texttt{specgroup}.

We fit the PN, MOS1, and MOS2 spectra in 0.3--10\,keV simultaneously with all the parameters tied between instruments. To further correct for any differences in calibrations between the detectors, we added an overall normalization constant for the MOS1 and MOS2 spectra. The discrepancies in the following analysis are well below $\lesssim$5\%. The X1 observation was taken in the tail of the canonical outburst, roughly five months after the \chandra\ C3 and C4 observations. The spectra can be well described with an MCD model. The best-fit $kT_{\rm{in}}$ is $0.79\pm0.06$\,keV and the apparent inner radius of the accretion disk is $R_{in}\sqrt{\cos{\theta}}=53_{-8}^{+9}$\,km, consistent with the C2 spectrum taken in 2011. Similar to all the \chandra\ observations, the fitting cannot be significantly improved by two-component or more complex models. The spectral change from C2 to X1 suggests that P9 had a canonical outburst in 2011--2012, and its spectrum switched between an MCD with an inner radius of $\sim50$\,km and a PL with $\Gamma\sim2.3$. 

\begin{deluxetable*}{cccccccccc}
\tabletypesize{\footnotesize}
\tablecaption{Best-fit Spectral Parameters of NGC~7793~P9 obtained from the \chandra, \xmm, and \nustar\ Data Sets. } 
\label{xmm_chandra_parameter}
\tablehead{\colhead{Obs.} & \colhead{Model} & \colhead{$L_{\textrm{0.3--10 keV}}$} & $L_{\rm{MCD}}$ & \colhead{$N_{\rm{H}}$} &\colhead{$R_{\rm{in}}\sqrt{\cos\theta}$} &\colhead{$kT_{\rm{in}}$} & \colhead{$\Gamma$} & \colhead{$\chi^2/\rm{dof}$} & \colhead{State} \\ 
 & & \colhead{($10^{39}$\,\lumcgs)} & \colhead{($10^{39}$\,\lumcgs)} & \colhead{($10^{22}$\,cm$^{-2}$)} & \colhead{(km)} & \colhead{(keV)} &  &  & }
\startdata
C1 &  PL & $0.10_{-0.02}^{+0.03}$ & \nodata & $0.13_{-0.13}^{+0.17}$ & \nodata & \nodata & $2.4_{-0.4}^{+0.5}$ & $6.8/14$ & SPL \\
C2 &  MCD & $0.5\pm0.1$ & $0.5\pm0.1$ & $0.07\pm0.04$ & $49^{+7}_{-6}$ & $0.96^{+0.07}_{-0.06}$  & \nodata & $62.7/57$ & Thermal \\
C3 &  PL & $0.79\pm0.06$ & \nodata & $0.09\pm0.04$ & \nodata & \nodata  & $2.2\pm0.1$ & $62.2/67$ & SPL \\
C4 & PL & $0.83_{-0.08}^{+0.10}$ & \nodata & $0.14_{-0.07}^{+0.08}$ & \nodata & \nodata  & $2.3\pm0.02$ & $27.1/31$ & SPL \\
X1 &  MCD & $0.26_{-0.07}^{+0.08}$ & $0.26_{-0.07}^{+0.08}$ & $0.024_{-0.014}^{+0.018}$ & $53^{+8}_{-7}$ & $0.78\pm0.02$  & \nodata & $77.8/93$& Thermal \\
X3 &  MCD+PL & $1.1\pm0.2$ & $0.36^{+0.07}_{-0.08}$ & $0.07\pm0.02$ & $370^{+90}_{-60}$ & $0.34\pm0.04$  & $1.8\pm0.2$ & $171.9/189$ & UL (HUL) \\
X4$^a$ &  MCD+PL & $3.4_{-0.4}^{+0.6}$ & $1.4^{+0.5}_{-0.4}$ & $0.11_{-0.02}^{+0.03}$ & $34^{+7}_{-6}$ & $1.5\pm0.2$  & $2.0_{-0.1}^{+0.2}$ & $308.3/311$ & UL (BD?)\\
 & MCD+PL$^b$ &  $4_{-2}^{+5}$ & $0.6_{-0.4}^{+0.8}$ & $0.06\pm0.04$ & $100_{-60}^{+90}$ & $0.8_{-0.2}^{+0.5}$ & $1.4_{-0.8}^{+0.4}$ & $302.1/309$ & \\
\enddata
\tablenotetext{a}{The \nustar\ data set was also included in the fitting.}
\tablenotetext{b}{A high-energy cutoff was included in this PL component with a cutoff energy of $4_{-4}^{+1}$\,keV and the folding energy of $5\pm5$\,keV.}
\vspace{-0.9cm}
\end{deluxetable*}

The X3 observation was made in the early phase of the ultraluminous outburst. We found that neither the MCD model nor the PL model provides acceptable fitting. On the other hand, a two-component MCD+PL model can significantly improve the fitting with a $\chi^2/dof=171.9/189$ (see Figure \ref{chandra_xmm_spec}). The best-fit $N_{\rm{H}}$ is $(7\pm2)\times10^{20}$\,cm$^{-2}$. The temperature and the radius of the inner disk are $kT_{\rm{in}}=0.34\pm0.04$\,keV and $R_{\rm{in}}\sqrt{\cos\theta}=370^{+90}_{-60}$\,km, which imply a much cooler disk with an innermost radius much farther away from the central compact object compared to the canonical thermal and SPL states of a BHXB. The PL component has $\Gamma=1.8(2)$, clearly harder than the value of the canonical SPL spectrum. 

The X4 observation was made near the peak of the ultraluminous outburst. We included the \nustar\ 3--20\,keV data in the fitting to increase the statistic and constrain the spectral behavior in the hard X-ray band. Since the \xmm\ X4 observation was made before this drop, we extracted the \nustar\ spectra during the X4 observation time interval.  We fit all the five spectra (PN, MOS1, MOS2, FPMA, FPMB) simultaneously. Similar to the X3 observation, a two-component MCD+PL model provides a good statistic with $\chi^2/\rm{dof}=308.3/311$. The disk temperature is $kT_{\rm{in}}=1.5\pm0.2$\,keV and $\Gamma=2.0\pm0.2$. In the soft X-ray band, the apparent spectral curvature caused by a highly optically thick Comptonized component may be classified as a disk-like spectrum \citep{SuttonRM2013}. Therefore, we used an MCD plus PL with a high-energy cutoff to fit the data. The fitting statistic is equally good with a $\chi^2/\rm{dof}=302.1/309$. The $N_{\rm{H}}$ value is $(7\pm4)\times10^{20}$\,cm$^{-2}$, consistent with that obtained in the previous section. The cutoff energy is $5\pm5$\,keV, indicating a strong spectral curvature caused by an optically thick corona but not well constrained. 

\subsubsection{Determining the Extragalactic Absorption}
Most of the best-fit models yield a similar $N_{\rm{H}}$ with a value of ($\sim$5--10)$\times10^{20}$\,cm$^{-2}$. Therefore, we assume that $N_{\rm{H}}$ did not significantly vary during the entire observation time span. We fit all the \chandra, \xmm, and \nustar\ data simultaneously with a tied $N_{\rm{H}}$ and their corresponding best-fit models to further constrain the $N_{\rm{H}}$. The resulting $\chi^2/\rm{dof}$ is $767.3/789$, and $N_{\rm{H}}$ is determined as $(7\pm1)\times10^{20}$\,cm$^{-2}$.  We also tried to use the abundances with the absorption model {\tt phabs} with the corresponding abundance and obtained $N_{\rm{H}}=(6\pm1)\times 10^{20}$\,cm$^{-2}$, slightly lower than that obtained with \texttt{tbnew} \citep{AndersG1989}. Other parameters are fully consistent with those listed in Table \ref{xmm_chandra_parameter}. The $N_{\rm{H}}$ is consistent with that of P13, indicating that P9 is unlikely a heavily absorbed source compared to other point sources in NGC 7793 \citep{IsraelPE2016}.

\subsection{More Physical Models}
From the spectral shapes in Figure \ref{chandra_xmm_spec} and fitting parameters in Table \ref{xmm_chandra_parameter}, we noticed that the X3 and X4 spectra likely represent the ultraluminous states. X3 can be well described with a cool disk plus a hard PL without a significant high-energy cutoff below 10\,keV. The spectral shape is similar to BHXB spectra in the low/hard state but the luminosity is much higher. This is the typical spectral feature of hard ULXs. The low $kT_{\rm{in}}<0.5$\,keV and hard $\Gamma<2$ implies an HUL spectrum according to the empirical classification \citep{SuttonRM2013}. The X4 spectrum is disk-like but much broader than a standard MCD. This spectrum can be classified as a BD. However, X4 shows a significant difference from X3 only above $\sim$1\,keV, implying that the major difference could be contributed by the PL tail. This heavily Comptonized spectra are sometimes misclassified as a BD. Hence, we carefully examine the X3 and X4 spectra with more physical Comptonized models and slim disk models to explore the underlying physics.

\subsubsection{Slim Disk}
We first fit X4 with a $p$-free accretion disk (DiskPBB), in which the radial advection is taken into account \citep{MineshigeHK1994, Hirano1995}. This model is an approximation of the slim disk. The radial temperature profile in this model has a PL dependence $T(r)\propto r^{-p}$, where $r$ is the radius and $p$ is a free parameter. A standard MCD corresponds to $p=0.75$, and a slim disk can be approximated with $p<0.75$. The best-fit $\chi^2/\rm{dof}$ is $308.3/312$, slightly better than the statistic of the MCD+PL model. The inner-disk temperature is $kT_{\rm{in}}=2.2\pm0.2$\,keV, much higher than that in the high/soft state of the normal outburst (see Table \ref{optxagnf_result}). The low $p=0.57\pm0.02$ is consistent with other ULXs with a BD spectrum \citep{KaaretFR2017}. The apparent inner radius, $R_{\rm{in}}\sqrt{\cos\theta}=16\pm3$\,km, is extremely small. We further tried a table model for a slim disk \citep{Kawaguchi2003}, in which included corrections for local BB contribution, Comptonized spectrum, or relativistic effects. We found that the spectrum can be described by the model ID 4, which includes a slim disk and a Comptonized local spectrum. The resulting $\chi^2/\rm{dof}$ is $304.7/311$, which is equally good as the $p$-free disk model. The BH mass is $4^{+27}_{-2}M_\odot$, $N_{\rm{H}}$ is $(7\pm2)\times10^{20}$\,cm$^{-2}$, the mass accretion rate is $\dot{m}=6^{+5}_{-1}$\,$L_{\rm{Edd}}/c^2$, but the viscosity is not well constrained. $N_{\rm{H}}$ is fully consistent with previous measurements. We also tried to use the $p$-free disk model to fit X3 spectra; however, the statistic is unacceptably high since the slim disk is unable to describe the high-energy tail. We therefore add a PL component and found that the disk component has a small value of $p$ but the uncertainty is large. 

\subsubsection{Comptonized Model: SIMPL}
The disk corona usually contributes a high-energy tail in the X-ray spectrum of a BHXB.  We described the tail with the convolution model SIMPL. This model empirically calculates the Comptonized spectrum from the scattering of a fraction of seed photons, i.e., the MCD component \citep{Steiner2009}. The seed photons are redistributed to both higher and lower energies. The normalization factor is the scattered fraction ($f_{\rm{sc}}$) We fit the X3 data set, which results in an acceptable statistic of $\chi^2/dof=172.9/189$. The disk is partially scattered with a fraction of $f_{\rm{sc}}=0.33$, and has an apparent inner-disk temperature of $kT_{\rm{in}}=0.33\pm0.04$\,keV and a radius of $R_{\rm{in}}\sqrt{\cos\theta}=500_{-100}^{+160}$\,km. This result still shows a truncated disk similar to that obtained from phenomenological models.

We fit the X4 data with the same model, but the best-fit statistic ($\chi^2/\rm{dof}=366.2/315$) is much worse than that yielded by MCD+PL. We obtained $f_{\rm{sc}}=1$, implying that all the disk photons are scattered to the Comptonized component. The seed MCD component has $kT_{\rm{in}}=0.92\pm0.6$\,keV and $R_{\rm{in}}\sqrt{\cos\theta}=109_{-12}^{+14}$\,km, a bit closer to the BH than that obtained from the X3 data set.

\subsubsection{Comptonized Model: NTHCOMP}
In addition to the convolution model, we used an additive Comptonized model, NTHCOMP, to describe the Comptonized continuum \citep{ZdziarskiJM1996, ZyckiDS1999}. The high-energy rollover, which is sharper than an exponential cutoff, is parameterized by the electron temperature $kT_e$. We set the maximum seed photon temperature $kT_{bb}$ to the inner-disk temperature $kT_{\rm{in}}$ of the MCD component. The statistic of X3 is insensitive to the electron temperature $kT_e$ because no significant high-energy cutoff is observed below 10\,keV. Hence, we fixed it to 100\,keV. The fit statistic is acceptable and the parameter suggests a cool disk with a large inner radius (see Table \ref{optxagnf_result}). The true disk radius can be re-estimated using the equation provided by \citet{KubotaM2004} for the presence of disk Comptonization. Assuming an inclination angle of 60$^{\circ}$, $R_{\rm{in}}$ can be estimated as $770_{-140}^{+300}$\,km, slightly larger than that directly obtained from the MCD component.

The Comptonized component of X4 can also be described with this model. However, the normalization of the disk component is consistent with zero, implying that MCD is negligible. Therefore, we used NTHCOMP to fit the spectra and set $kT_{bb}$ as a free parameter. The statistic is as good as that obtained from phenomenological models. The best-fit $kT_e=1.9_{-0.3}^{+0.5}$, roughly corresponds to an exponential cutoff at 4--7\,keV. This is consistent with that obtained from the MCD+PL model with a high-energy cutoff. The inner-disk radius can be re-estimated as $460_{-140}^{+280}$\,km, consistent with that obtained from the X3 data set.

\begin{figure*}
\begin{minipage}{0.49\linewidth}
\includegraphics[width=0.95\textwidth]{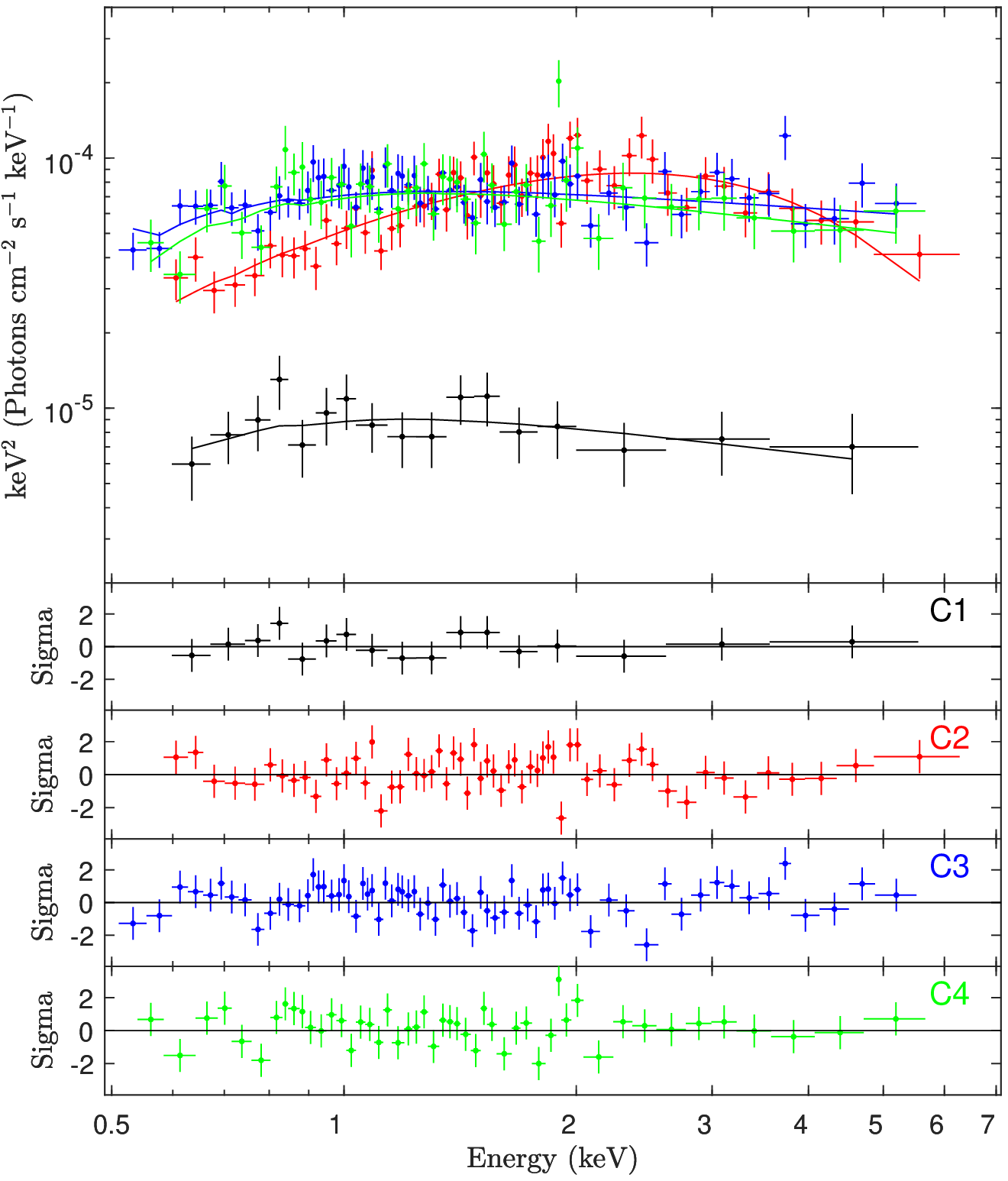}
\end{minipage}
\begin{minipage}{0.49\linewidth}
\includegraphics[width=0.95\textwidth]{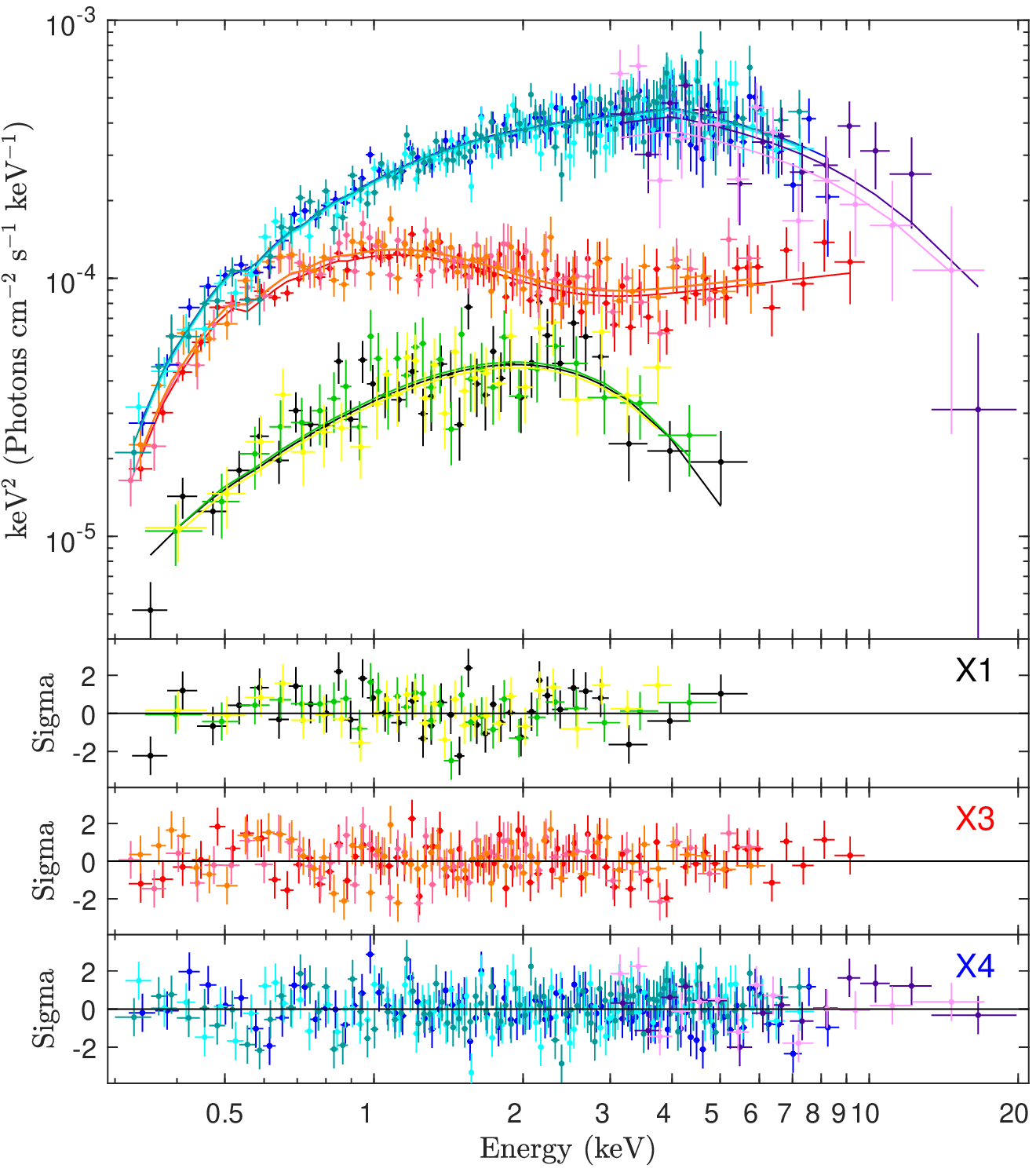}
\end{minipage}
\caption{(Left) The \chandra\ spectra of NGC~7793~P9. The black, red, blue, and green points represent the C1, C2, C3, and C4 spectra, respectively. The solid lines are the corresponding best-fit models. (Right) The \xmm\ and \nustar\ spectra of NGC~7793~P9. The solid lines are the corresponding best-fit MCD, MCD+PL, and DiskPBB models for the X1, X3, and X4 data sets, respectively.}
\label{chandra_xmm_spec}
\end{figure*}

\subsubsection{Disk--Corona Coupling: OPTXAGNF}
The above two models account for the partial Comptonization of a standard MCD. However, the size of the corona may play an important role if the disk is only partially covered. The innermost accretion disk could be fully or partially Comptonized, while the outer disk remains standard and contributes a significant amount of soft X-ray emission.  \citet{DoneK2006} proposed the DKBBFTH model to deal with the spectrum from the Comptonized corona over the inner disk. This model has been applied to the spectra of several ULXs to recover the properties of the inner disk \citep{GladstoneRD2009}. An upgraded model, OPTXAGNF, was proposed to fit the broadband spectra of narrow-line Seyfert 1 galaxies with the same aspect \citep{DoneDJ2012}. Given a similar accretion property, this model can be used to describe the UL and SPL states of BHXBs. The disk component is a canonical MCD where the radius is larger than the coronal radius. The disk inside the corona is not fully thermalized. Instead, the emission was Comptonized to form both soft excess and a hard PL tail. There are 11 parameters in OPTXAGNF: the BH mass ($M_{\rm{BH}}$), distance, luminosity in units of Eddington ratio ($\log L/L_{\rm{Edd}}$), BH spin ($a$), coronal radius ($R_{\rm{cor}}$ in units of $r_g=GM_{\rm{BH}}/c^2$), outer radius of the disk ($R_{\rm{out}}$), electron temperature ($kT_e$) and optical depth ($\tau$) for the soft Comptonization component, photon index of the hard Comptonization component ($\Gamma$), fraction ($f_{\rm{pl}}$) of the power below $R_{\rm{cor}}$ that is redistributed to the hard PL, and the redshift. We fit the X3 and X4 data sets simultaneously and tied the BH mass with each one, and assumed $N_{\rm{H}}=7\times10^{20}$\,cm$^{-2}$ and $a=0$. We found that the $R_{\rm{cor}}$ for these two data sets hit the hard limit. We therefore set $R_{\rm{cor}}=100$\,$r_g$. A BH mass of $M_{\rm{BH}}=22\pm3$\,$M_\odot$ can result in a good statistic (see Table \ref{optxagnf_result}). The luminosities are $1.1\pm0.2$\,$L_{\rm{Edd}}$ and $1.3\pm0.2$\,$L_{\rm{Edd}}$ for the X3 and X4 data sets, respectively. The Comptonized component has a high $\tau$ for both data sets, implying an optically thick corona. We also tried to set the BH spin to the maximum value of $a=0.998$. In this fitting, $R_{\rm{cor}}$ is well constrained but $\tau$ is not. The fitting is insensitive to $kT_e$. Hence, we fixed it to 1\,keV to account for the soft excess. We found that the BH mass is $120\pm20$\,$M_\odot$, six times larger than that of the $a=0$ case. The corona has a radius of $\sim$15\,$r_g$ for both the X3 and X4 spectra, roughly the same size as $\sim$100\,$r_g$ of a $\sim$20\,$M_{\odot}$ BH. Similar to the best-fit result for a non-spinning BH, a highly Comptonized, optically thick corona is required for both data sets. 

\subsection{Spectral Variability of \swift\ Observations}
\begin{figure}
\centering
\includegraphics[width=0.47\textwidth]{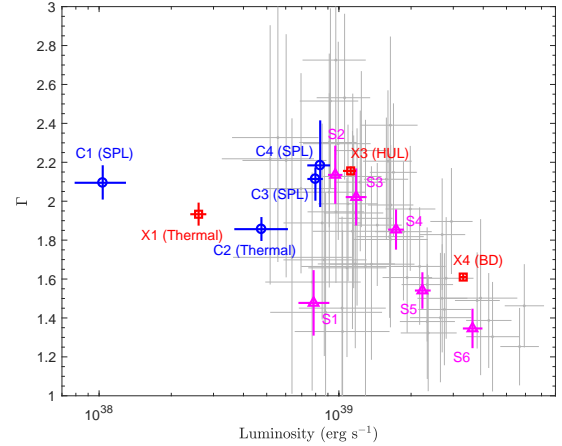} 
\caption{Photon indices obtained by all the \swift\ individual observations (gray), \chandra\ (blue circle), \xmm\ (red square), and stacked \swift\ observations (magenta triangle). }
 \label{pl_index_all}
\end{figure}

To obtain an overall hardness--intensity relationship, we fit the individual \swift, \chandra, and \xmm\ data sets with a simple PL model and fixed $N_{\rm{H}}$ to $7\times10^{20}$\,cm$^{-2}$ (see Figure \ref{pl_index_all}). From the overall $\Gamma$--$L$ relationship obtained from individual \swift\ observations, we found that $\Gamma$ varied over a wide range when the luminosity is lower than $10^{39}$\,\lumcgs, likely caused by the transition between the SPL and the high/soft states as seen in canonical BHXBs. In addition, $\Gamma$ shows a break and a decreasing trend when the luminosity is higher than $10^{39}$\,\lumcgs. We then stacked the \swift\ observations with similar luminosities and spectral behaviors to increase the signal-to-noise ratio (see Table \ref{swift_parameter} for definition and result). We fit the six data sets individually with $N_{\rm{H}}$ fixed to $7\times10^{20}$\,cm$^{-2}$ using the the Cash statistic. We also binned the spectra and fit them with the minimum $\chi^2$ algorithm, and the results were fully consistent with those obtained from Cash statistic. We fit the spectra with MCD+PL to get hints about the spectral behavior. We did not attempt to use more physical models to fit the \swift\ spectra due to a limited number of photons.

\begin{deluxetable}{llcc}
\tabletypesize{\footnotesize}
\tablecaption{Best-fit Spectral Parameters of the X3 and X4 Data Sets with More Physical Models. } 
\label{optxagnf_result}
\tablehead{\colhead{Model} & \colhead{Parameters} & \colhead{X3} & \colhead{X4}}
\startdata
 & $N_{\rm{H}}$ ($10^{22}$\,cm$^{-2}$) & $0.12\pm0.05$ & $0.09\pm0.02$ \\
 & $T_{\rm{in}}$\,(keV) & $0.5_{-0.1}^{+0.2}$ & $2.2\pm0.2$ \\
DiskPBB+PL& $R_{\rm{in}}\sqrt{\cos\theta}$\,(km) & $120_{-80}^{+190}$ & $16\pm3$\\
 & $p$ & $0.47_{-0.06}^{+0.20}$ & $0.57\pm0.02$ \\
 & $\Gamma$ & $1.5_{-0.5}^{+0.3}$ & \nodata\ \\
 & $\chi^2/\rm{dof}$ & $168.2/188$ & $308.3/312$ \\
\hline
 & $N_{\rm{H}}$ ($10^{22}$\,cm$^{-2}$) & $0.05\pm0.02$ & $0.007\pm0.007$ \\
 & $T_{\rm{in}}$\,(keV) & $0.33\pm0.04$ & $0.92_{-0.06}^{+0.07}$ \\
SIMPL(MCD)& $R_{\rm{in}}\sqrt{\cos\theta}$\,(km) & $500_{-100}^{+170}$ & $110_{-12}^{+14}$\\
 & $\Gamma$ & $1.8\pm0.2$ & $2.9_{-0.2}^{+0.3}$ \\
 & $f_{\rm{sc}}$ & $0.33_{-0.07}^{+1.0}$ & $1.0^a$ \\
 & $\chi^2/\rm{dof}$ & $172.9/189$ & $367.4/314$ \\
\hline
 & $N_{\rm{H}}$ ($10^{22}$\,cm$^{-2}$) & $0.05\pm0.02$ & $0.04_{-0.01}^{+0.02}$ \\
 & $T_{\rm{in}}$\,(keV) & $0.33\pm0.04$ & $0.4\pm0.1^b$ \\
 & $R_{\rm{in}}\sqrt{\cos\theta}$\,(km) & $400_{-70}^{+100}$ & \nodata\ \\
MCD+NTHCOMP & $kT_e^d$\,(keV) & $100^a$ & $1.9_{-0.3}^{+0.5}$ \\
 & $\Gamma$ & $1.8\pm 0.2$ & $1.9\pm 0.1$ \\
 & $\chi^2/\rm{dof}$ & $173.2/189$ & $310.7/313$ \\
\hline
 & BH Mass ($M_\odot$) &  \multicolumn{2}{c}{$22\pm3$} \\
 & $\log L$ ($L_{\rm{Edd}}$) & $0.02\pm0.07$ & $0.10\pm0.08$ \\
 & $R_{\rm{cor}}$ ($r_g$) & $100^a$ & $100^a$ \\
OPTXAGNF & $\tau$ & $100^a$ & $15_{-2}^{+7}$ \\
($a=0$) & $\Gamma$ & $1.60_{-0.05}^{+0.04}$ & $1.9\pm0.3$ \\
 & $kT_e^e$ (keV) & $0.5\pm0.1$ & $1.2\pm0.2$ \\
 & $f_{\rm{pl}}$ & $0.97\pm0.01$ & $0.4\pm0.2$\\ 
 & $\chi^2/\rm{dof}$ & \multicolumn{2}{c}{$500.6/502$}\\
\hline
 & BH Mass ($M_\odot$) &  \multicolumn{2}{c}{$120\pm20$} \\
 & $\log L$ ($L_{\rm{Edd}}$) & $-0.7_{-0.1}^{+0.2}$ & $-0.6_{-0.1}^{+0.2}$ \\
 & $R_{\rm{cor}}$ ($r_g$) & $15_{-2}^{+3}$ & $15_{-6}^{+12}$ \\
OPTXAGNF & $\tau$ & $24_{-7}^{+76\,c}$ & $100^a$ \\
($a=0.998$) & $\Gamma$ & $1.61 \pm 0.06$ & $2.0\pm0.2$ \\
 & $kT_e^e$ (keV) & $1.0^a$ & $1.0^a$ \\
 & $f_{\rm{pl}}$ & $1.0^a$ & $0.71_{-0.08}^{+0.01}$ \\
 & $\chi^2/\rm{dof}$ & \multicolumn{2}{c}{$510.6/503$}\\
\enddata
\tablenotetext{a}{This parameter is frozen.}
\tablenotetext{b}{This temperature is $kT_{bb}$ in the NTHCOMP model because the MCD component is not included.}
\tablenotetext{c}{The upper bound of $\tau$ is not well constrained.}
\tablenotetext{d}{The electron temperature in the NTHCOMP model denotes the high-energy rollover. An exponential cutoff could be roughly estimated as $E_c=$2--3\,$kT_e$.}
\tablenotetext{e}{The electron temperature in the OPTXAGNF model denotes the soft excess caused by Comptonization. The electron temperature for the hard Comptonization component is fixed to 100\,keV.}
\vspace{-0.9cm}
\end{deluxetable}

The S1 spectrum ($L<10^{39}$\,\lumcgs) is obviously disk-like. It can be well described by the MCD model with $kT_{\rm{in}}=1.3_{-0.2}^{+0.3}$\,keV and $R_{\rm{in}}\sqrt{\cos\theta}=33^{+11}_{-9}$\,km. We attempted to fit the S1 spectrum with the DiskPBB model and found $p=0.7_{-0.1}^{+0.2}$, fully consistent with that of an MCD. The S2--S4 data sets are better described by a simple PL compared to the MCD. We obtained an anticorrelation between $\Gamma$ and flux from these three data sets (see Figure \ref{pl_index_all}). We tried to fit the spectra with the MCD+PL model, and the statistics were slightly improved. The best-fit MCD component suggested that the disk is cool with $kT_{\rm{in}}\sim0.5$\,keV and the inner radius is larger than the innermost stable circular orbit. We tried the DiskPBB model and yielded equally good statistics. The best-fit $p$-values are much smaller than those obtained from X4 and close to the hard limit of the DiskPBB model. 

\begin{figure}
\includegraphics[width=0.47\textwidth]{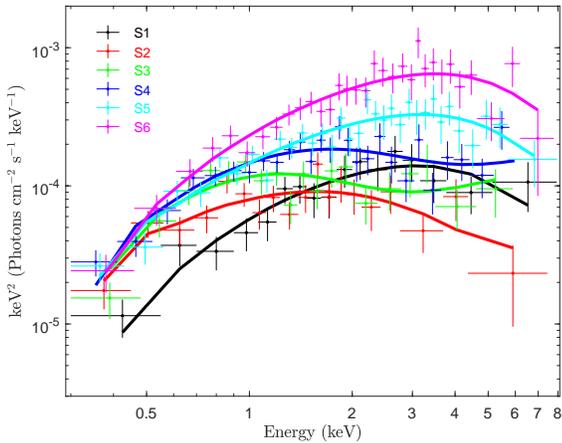} 
\caption{\swift\ stacked spectra and the best-fit MCD+PL model P9. }
\label{swift_spec_state}
\end{figure}

On the other hand, the S5 and S6 spectra are disk-like and can be well described with an MCD.  Adding a PL component to S5 slightly improved the statistic but the change in the MCD component is limited and the PL component was not well constrained. The MCD fit of S5 has $kT_{\rm{in}}=1.4_{-0.3}^{+0.4}$\,keV and $R_{\rm{in}}\sqrt{\cos\theta}=46^{+14}_{-13}$\,km, similar to the MCD component in X4. This spectrum is equally well described with a $p$-free slim disk. The best-fit $p=0.65_{-0.07}^{+0.07}$ is consistent with that obtained in the X4 data set but much closer to that of a standard MCD. For the S6 spectrum, adding a PL component or fitting $p$ could not improve the fit statistic. The parameters of the PL components are not constrained, and the $p$-value is fully consistent with 0.75.  We plot all of the stacked \swift\ spectra with the best-fit model in Figure \ref{swift_spec_state} and rebin the spectral channels to have at least 5$\sigma$ significance. It is clear that the spectral shape changes from PL-like to disk-like as flux increased when $L\gtrsim10^{39}$\,\lumcgs. S3 and S4 are similar to X3. They show a hint of a hard PL tail although the significance is low due to the limited energy range and low photon statistic. S5 and S6 are disk-like, but they do not have an obvious deviation from the S3 and S4 spectra below $\sim1$\,keV. This behavior is similar to that of X4, implying that the apparent disk-like spectrum could be a misleading.

\section{Discussion}\label{discussion}

\begin{deluxetable*}{ccccccccc}
\tabletypesize{\footnotesize}
\tablecaption{Best-fit Spectral Parameters of Stacked \swift\ Spectra During the Ultraluminous Outburst. } 
\label{swift_parameter}
\tablehead{\colhead{Bins} & \colhead{Luminosity Range} & \colhead{Model} & \colhead{$L_{\textrm{0.3--10 keV}}^a$} & \colhead{$L_{\textrm{MCD}}$} &\colhead{$R_{\rm{in}}\sqrt{\cos\theta}$} &\colhead{$kT_{\rm{in}}$} & \colhead{$\Gamma$ or $p$} & \colhead{$C^2/\rm{dof}$} \\ 
\colhead{} & \colhead{($10^{39}$\,\lumcgs)} & \colhead{} & \colhead{($10^{39}$\,\lumcgs)} & \colhead{($10^{39}$\,\lumcgs)} & \colhead{(km)} & \colhead{(keV)} & \colhead{} & \colhead{} }
\startdata
S1 & $L<1$, $\Gamma<1.8$ & MCD & $0.8\pm0.1$ & $0.8\pm0.1$ & $33^{+11}_{-9}$ & $1.3^{+0.3}_{-0.2}$ & \nodata & $143.9/167$  \\
S2 & $L<1$, $\Gamma>1.8$ & MCD+PL & $0.86\pm0.8$ & $0.3\pm0.1$ & $150_{-100}^{+120}$ & $0.5_{-0.1}^{+0.3}$ & $1.9(5)$ & $224.8/254$ \\
 &  & DiskPBB &  & \nodata & $<48$ & $1.1_{-0.2}^{+0.4}$ & $0.53_{-0.03}^{+0.04\,b}$ & $223.2/255$ \\ 
S3 & $L=1$--1.3 & MCD+PL & $1.2\pm0.8$ & $0.4\pm0.2$ & $300_{-150}^{+180}$ & $0.39_{-0.08}^{+0.12}$ & $1.4_{-0.9}^{+0.5}$ & $149.1/184$ \\
 & & DiskPBB & & \nodata & $<25$ & $1.3_{-0.4}^{+1.5}$ & $0.53_{-0.04}^{+0.05\,b}$ & $154.3/184$ \\
S4 & $L=1.3$--2 & MCD+PL & $1.5\pm0.2$ & $0.6_{-0.3}^{+0.5}$ & $150_{-80}^{+90}$ & $0.6_{-0.1}^{+0.2}$ & $1.5_{-0.9}^{+0.4}$ & $258.3/267$ \\
 & & DiskPBB & & \nodata & $23_{-21}^{+18}$ & $1.4_{-0.3}^{+0.8}$ & $0.56_{-0.03}^{+0.04\,b}$ & $264.8/267$ \\
S5 & $L=2$--3 & MCD+PL & $2.2\pm0.2$ & $1.7_{-0.6}^{+0.3}$ & $46^{+14}_{-13}$ & $1.4^{+0.2}_{-0.3}$ & $2.3^{+6.2}_{-1.2}$ & $280.0/324$ \\
 & & DiskPBB &  & \nodata & $34^{+20}_{-17}$ & $1.5^{+0.4}_{-0.2}$ & $0.65^{+0.07\,b}_{-0.05}$ & $279.8/325$ \\
S6 & $L>3$ & MCD & $3.6_{-0.3}^{+0.4}$ & $3.6_{-0.3}^{+0.4}$ & $56^{+11}_{-10}$ & $1.5^{+0.2}_{-0.1}$ & \nodata & $253.5/312$ 
\enddata
\tablenotetext{a}{Derived from the best-fit model.}
\tablenotetext{b}{This is the $p$-value of the $p$-free disk.}

\vspace{-0.9cm}
\end{deluxetable*}

\subsection{Implications from the MCD+PL Model}
\subsubsection{Spectral Evolution of P9}
During the canonical outburst, the X-ray spectrum of P9 switched between an MCD and a PL. The MCD spectra were observed in mid 2011 and mid 2012. The disk had an apparent inner radius of $R_{\rm{in}}\sqrt{\cos\theta}\approx50$\,km and an inner-disk temperature $kT_{\rm{in}}=0.7$--0.9\,keV. This radius corresponds to a true inner radius of 90\,km after applying corrections for the color–temperature and inner boundary condition of an MCD and assuming an inclination angle of 60$^{\circ}$ \citep{ShimuraT1995, KubotaTM1998}. Assuming a non-spinning BH, the mass can be estimated to be $\sim10$\,$M_\odot$. For an extremely spinning BH, the apparent value corresponds to a BH mass of $\sim60$\,$M_\odot$. In addition, the disk could be weakly Comptonized and result in a smaller $R_{\rm{in}}$ \citep{KubotaM2004}. The above mass estimation represents the lower limit of the BH mass in P9. This indicates that P9 has a regular or massive stellar-mass BH.  Its X-ray spectrum switched between the SPL and high/soft states of a canonical outburst before 2013 \citep{TetarenkoSH2016}.

When the luminosity exceeds $10^{39}$\,\lumcgs, $\Gamma$ shows a clear decreasing trend with the luminosity (see Figure \ref{pl_index_all}), and the high-quality \xmm\ spectra in this ultraluminous state cannot be well fit by single-component models. They are significantly different from those in the canonical outburst. When the luminosity is slightly higher than $L\approx10^{39}$\,\lumcgs\ (X3, S3, and S4), P9 has a cool disk and a hard PL tail. Moreover, the spectral curvature increases with flux when $L\gtrsim2\times10^{39}$\,\lumcgs\ (X4, S5, S6). These data sets can be described by disk-like spectral models.

As described in Section \ref{introduction}, the BD usually dominates the spectra of ULXs with $L\lesssim3\times10^{39}$\,\lumcgs, and those of ULXs with $L\gtrsim3\times10^{39}$\,\lumcgs\ showed two-component spectra like HUL and SUL \citep{GladstoneRD2009, SuttonRM2013}. However, some ULXs like NGC 1313 X-2 and Holmberg IX X-1 showed an inverse transition, i.e., an HUL spectrum in the low-luminosity state, and a disk-like BD spectrum near their peak luminosities \citep{PintoreZ2012, MiddletonHP2015, LuangtipRD2016}. The BD feature of these sources is likely to be misclassified due to the strong spectral curvature caused by the highly optically thick Comptonized spectrum \citep{SuttonRM2013}. Similar to these two sources, P9 has an apparent inverse transition between these two states. However, the X4 light curve shows a significant variability dominated by the emission above 1~keV ($F_{\rm{var}}=0.15(2)$ for 1--10\,keV and $F_{\rm{var}}=0.05(2)$ for 0.3--1\,keV). It is difficult to interpret with a single slim disk model since the variability should be independent of the energy for a single emission component. Hence, we suggest that the apparent BD spectral shape observed in X4 consists of two components like NGC 1313 X-2 and Holmberg IX X-1.

\subsubsection{Variability of the Disk Component}
P9 evolved from the canonical BHXB regime to the ULX regime when its luminosity crossed the Eddington limit of a $\sim10$\,\ms\ BH. To obtain the disk evolution, we calculated the 0.3--10\,keV luminosity of the disk to check its correlation with the inner-disk temperature. Assuming a fixed inner radius, the luminosity of a standard disk shows a positive correlation with the disk temperature:
\begin{equation}
L_{\rm{MCD}}=\frac{\pi\sigma G^2M^2T_{\rm{in}}^4}{6c_0^4f_{\rm{col}}^4c^4},
\end{equation}
where $\sigma$ is the Stefan--Boltzmann constant, $M$ is the mass of the central compact object, $c_0\approx0.1067$, and $f_{\rm{col}}^4$ is the color--temperature correction, which is usually larger than unity \citep{GierlinskiZP1999, GierlinskiD2004}. Many ULXs (especially supersoft ones) show an anticorrelation between the disk luminosity and the temperature \citep[see, e.g.,][]{KajavaP2009, UrquhartS2016}. On the other hand, some ULXs show a positive correlation close to $L_{\rm{disk}}\propto T^4$ \citep{MillerFM2003}. Since P9 has both canonical and ultraluminous outbursts, it provides a unique chance to test the accretion disk model. 

\begin{figure}
\begin{center}
\includegraphics[width=0.5\textwidth]{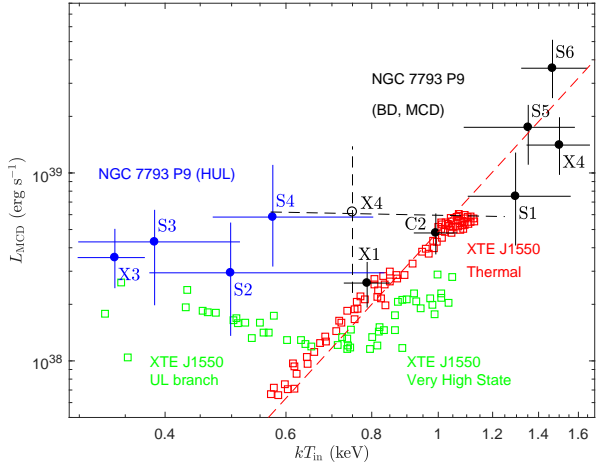} 
\end{center}
\caption{Comparison of the $kT_{\rm{in}}$--$L_{\rm{MCD}}$ relation between P9 and XTE J1550$-$564 \citep{Soria2007}. The red dashed line is the best-fit $L_{\rm{MCD}}\propto T_{\rm{in}}^4$ relation for XTE J1550$-$564. The red squares are obtained from the high/soft state of XTE J1550$-$564 and the green ones are from the SPL state and ultraluminous branch. The black data points denote the thermal and BD states of P9, and the blue ones denote the HUL spectral state. The black circle with the dashed line is the best-fit thermal component of X4 spectra with a two-component model containing an MCD and a PL with a high-energy cutoff. }
\label{kt_lt_1550}
\end{figure}

We collected the 0.3--10\,keV luminosity of the MCD component and disk temperature from both the high/soft state (C2, X1, and S1), HUL state (X3, S3, S4, and possibly S2), and disk-like state in the ultraluminous outburst (X4, S5, and S6). We show the disk luminosity--temperature relation of P9 together with that of a BHXB, XTE J1550$-$564, in Figure \ref{kt_lt_1550}. The disk configuration of P9 during the high/soft state in the canonical outburst (X1 and C2) was fully consistent with the thermal track of XTE J1550$-$564. We  attempted to add an NTHCOMP component to these two data sets but the Componized component contributed negligible emission. The disk component in the disk-like spectral states of P9 during the ultraluminous outburst well lies on the extension of the thermal track. This is likely a coincidence since they can also be well described by Comptonized models. The MCD component in the HUL state clustering at the lower-left corner of this figure is similar to the ultraluminous branch of XTE J1550$-$564. However, P9 shows a much harder ($\Gamma\sim$1.6--1.8) PL tail compared to XTE J1550$-$564 ($\Gamma\sim$2.5--2.8). The presence of this state and the apparent disk-like state makes P9 as a unique source among canonical BHXBs.

\subsection{Implications from the Comptonized Disk}
The simple MCD+PL model is helpful for classifying the spectral states. The result suggests that the innermost region of the accretion disk was covered by the Comptonized corona. We further used physical models to deal with the coupling of both the disk and the corona. We noticed that all of the spectra with luminosities above $L\gtrsim10^{39}$\,\lumcgs, including X3, X4, and S3--S6, show clear spectral difference above $\sim$1\,keV (see Figures \ref{chandra_xmm_spec} and \ref{swift_spec_state}). This implies that the mass accretion rate during the ultraluminous outburst has no significant difference, and the observed variability over (1--5)$\times10^{39}$\,\lumcgs\ is dominated by the change of the Comptonized component. If we use NTHCOMP to describe the Comptonized corona, then a cool inner-disk temperature, a large re-estimated radius, a hard PL with $\Gamma\sim1.8$, and a large optical depth are necessary for both X3 and X4 spectra. In contrast, X4 shows a higher normalization and a clear cutoff at 4--6\,keV. This implies that the plasma temperature of the corona is hot at the beginning of the ultraluminous outburst, and a huge amount of disk photons are scattered to hard X-ray. As the corona temperature cools down, fewer photons can be scattered to hard X-rays. The spectrum shows a cutoff in the \xmm\ energy range and an apparently higher luminosity in 0.3--10\,keV.

We also described the UL spectra with the OPTXAGNF model.  The BH mass can be estimated as $\sim20$\,$M_\odot$ for a non-spinning BH or $\sim120$\,$M_\odot$ for an extremely fast spinning BH. This result is a bit higher than that obtained from the simple estimation based on the apparent inner-disk radius in the high/soft state. In the case of $M_{\textrm{BH}}\sim20M_\odot$, most of the disk emission below $R_{\rm{cor}}$ is scattered into the hard Comptonized component ($f_{\rm{pl}}\sim1$) and produces a hard PL with $\Gamma=1.6$ in X3 spectrum. When the X-ray luminosity increases to near the peak of the ultraluminous outburst, a significant power is emitted in the low-temperature soft Comptonization component. This result is consistent with the cooling of the corona, which is implied from the NTHCOMP model. The luminosity is close to or slightly higher than the Eddington luminosity, which suggests that P9 belongs to the major ULX population: the high-luminosity tail of BHXBs. Of course, we could not exclude the explanation that X4 shows a slim disk since the spectrum can be well fit by both the DiskPBB model and the slim disk model. If this is true, the corona just fades away when the luminosity is close to the peak. However, the short-term variability is only observed above 1\,keV in the X4 observation. Therefore, the apparent BD spectrum likely contains two components.

A BH with a mass of $\sim120M_\odot$ can also produce the observed spectra with a smaller coronal radius (in units of $r_g$). The luminosity is roughly 0.2\,$L_{\rm{Edd}}$, consistent with that of canonical outbursts in BHXBs. If this is true, P9 is consistent with a massive stellar-mass BH that was recently found via gravitational wave observations \citep{AbbottAA2016, AbbottAA2017}. However, the hard PL tail observed in X3 is not expected in canonical outbursts. Hence, this scenario is less favored. 

\subsection{Supercritical Accretion}
The UL spectra can also be physically interpreted with the supercritical accretion model. The inner disk is geometrically thick and an optically thick outflowing wind is blown. A funnel-shaped cavity forms along the rotational axis of the BH \citep{KingDW2001, OhsugaM2007, MiddletonHP2015}.  In this case, the soft emission component is contributed by the wind, and the hard emission component is from the inner accretion disk or the accretion flow, although detailed modeling is still under development \citep{PoutanenLF2007, KajavaPF2012}.  The beaming factor was proportional to the mass accretion rate. Hence, the inclination angle is expected to be low since the hard component dominates the observed flux during the HUL state. The mass accretion rate is maximum during the HUL state (X3), where the beaming factor is large and we observed the innermost accretion flow.  When the mass accretion rate decreased, the inner-disk temperature slightly cooled down and the beaming factor decreased. The resulting spectrum is best described by a BD if the emission from the wind becomes less dominant.  As the disk luminosity drops to sub-Eddington levels, we observe a thin disk in the tail of the outburst. We attempted to fit the hard component with DiskPBB in the X3 data set, but the disk temperature is extremely high ($kT\sim4$\,keV) and the disk radius is not well constrained. Therefore, this scenario is less favored although not entirely excluded. Of course, a corona can possibly form around the supercritical accretion disk and cause a much more complex X-ray spectrum. 

Finally, P9 shows the HUL spectrum when $L\approx10^{39}$\,\lumcgs, much lower than several genuine ULXs showing the same spectral behavior \citep[e.g., 1--1.2$\times10^{40}$\lumcgs\ for Holmberg IX X-1, ][]{SuttonRM2013}. This could imply that those genuine ULXs have much higher BH masses if the spectral state classification only depends on the Eddington ratio. However, the luminosity of a ULX can be strongly affected by the beaming factor and viewing angle. An important counterexample is NGC 5907 ULX-1, which shows an HUL spectral behavior at $L\gtrsim10^{40}$\,\lumcgs. Recently, it was found to be powered by a neutron star \citep{IsraelBS2017}. Therefore, the spectral state and the mass of the central compact object cannot be directly associated. The empirical luminosity range of the spectral states may stop being valid as the ULX sample grows. Comprehensively analyzing the timing and spectral behaviors of ULXs with next-generation X-ray telescopes and improving the supercritical accretion spectral model will be helpful for describing the accretion disk structure of ULXs.

\section{Summary}\label{summary}
We carried out a systematic analysis of the spectral variability of the newly discovered transient ULX NGC 7793 P9 with \swift\ XRT monitoring and dedicated \chandra, \xmm, and \nustar\ observations. P9 is an LMXB and showed a canonical outburst behavior before 2013. The ultraluminous outburst after 2014 suggests that a BHXB can evolve to a ULX, supporting the scenario that some ULXs are powered by long-lasting super-Eddington outbursts. P9 is an important example bridging the BHXB and the ULX and shows the most fruitful spectral states among all the ULXs. The spectral analysis implies that the compact object is a stellar-mass BH. The simple MCD+PL model suggests that P9 shows an ultraluminous branch, similar to XTE~J1550$-$564. However, the spectrum is dominated by a much harder PL compared to XTE~J1550$-$564. This is the feature of an HUL spectrum of a ULX. We suggest that the inner accretion disk is covered by a corona that scatters most of the inner-disk emission to a hard PL. In addition, P9 evolved to an apparent BD spectral state when $L\gtrsim3\times10^{39}$\,\lumcgs\ although it is likely misclassified due to the strong curvature of the Comptonized component. We suggested that the corona cools down but the disk component remains unchanged. The soft Comptonization dominated the observed spectrum. The spectral evolution can also be interpreted with the supercritical accretion model, although a detailed spectrum of this model is still under development. The theoretical models can be further tested by tracking the spectral evolution of more samples of transient ULXs.

\acknowledgments

We thank Prof. Chris Done for useful suggestions in modeling the X-ray spectra and classifying the spectral states. We thank the anonymous referee for the comments that improved this paper. This research is in part based on the data obtained from the Chandra Data Archive, and has made use of software provided by the Chandra X-ray Center (CXC) in the application packages CIAO, ChIPS, and Sherpa. This research has used the observations obtained with \emph{XMM-Newton} and ESA science mission with instruments and contributions directly funded by the ESA member states and NASA. This research has also made use of data obtained with \emph{NuSTAR}, a project led by Caltech, funded by NASA and managed by NASA/JPL, and has utilized the NUSTARDAS software package, jointly developed by the ASDC (Italy) and Caltech (USA). This work made use of data supplied by the UK Swift Science Data Centre at the University of Leicester. C.P.H.~and C.Y.N.~are supported by a GRF grant from the Hong Kong Government under HKU 17300215P. A.K.H.K.~is supported by the Ministry of Science and Technology of the Republic of China (Taiwan) through grants 103-2628-M-007-003-MY3 and 105-2112-M-007-033-MY2.

\facilities{\emph{CXO} (ACIS), \emph{XMM} (MOS, PN), \emph{NuSTAR}, \swift\ (XRT, UVOT)}


\end{document}